\documentclass[pra,aps,twocolumn,nopacs,superscriptaddress,nofootinbib]{revtex4}

\usepackage{graphicx}  
\usepackage{dcolumn}  
\usepackage{bm}           
\usepackage{amsmath}
\usepackage{epsfig}
\usepackage{indentfirst}
\usepackage{psfrag}
\usepackage{subfigure}
\usepackage{amssymb}
\usepackage{color}
\usepackage[colorlinks,linkcolor=blue,citecolor=blue,urlcolor=blue,hyperindex,driverfallback=dvipdfm]{hyperref}
\usepackage[T1]{fontenc}
\def\ii{{\rm i}}  \def\ee{{\rm e}}

  \def\kB{{k_{\rm B}}}

\def\vF{v_{\rm F}}    \def\EF{{E_{\rm F}}}
        \def\wp{\omega_{\rm p}}
\def\eh{\epsilon_{\rm h}}  \def\Te{T_{\rm e}}  \def\Tl{T_{\rm l}} \def\ke{\kappa_{\rm e}}  \def\kl{\kappa_{\rm l}}
\def\Iinc{I^{\rm inc}}  

\begin{document}
\title{Photothermal Engineering of Graphene Plasmons}
\author{Renwen~Yu}
\affiliation{ICFO-Institut de Ciencies Fotoniques, The Barcelona Institute of Science and Technology, 08860 Castelldefels (Barcelona), Spain}
\author{Qiushi~Guo}
\affiliation{Department of Electrical Engineering, Yale University, New Haven, Connecticut 06511, USA}
\author{Fengnian~Xia}
\affiliation{Department of Electrical Engineering, Yale University, New Haven, Connecticut 06511, USA}
\author{F.~Javier~Garc\'{\i}a~de~Abajo}
\email[Corresponding author: ]{javier.garciadeabajo@nanophotonics.es}
\affiliation{ICFO-Institut de Ciencies Fotoniques, The Barcelona Institute of Science and Technology, 08860 Castelldefels (Barcelona), Spain}
\affiliation{ICREA-Instituci\'o Catalana de Recerca i Estudis Avan\c{c}ats, Passeig Llu\'{\i}s Companys 23, 08010 Barcelona, Spain}

\begin{abstract}
Nanoscale photothermal sources find important applications in theranostics, imaging, and catalysis. In this context, graphene offers a unique suite of optical, electrical, and thermal properties, which we exploit to show self-consistent active photothermal modulation of its nanoscale response. In particular, we predict the existence of plasmons confined to the optical landscape tailored by continuous-wave external-light pumping of homogeneous graphene. This result relies on the high electron temperatures achievable in optically pumped clean graphene while its lattice remains near ambient temperature. Our study opens a new avenue toward the active optical control of the nanophotonic response in graphene with potential application in photothermal devices.
\end{abstract}
\date{\today}
\maketitle

\section{Introduction}

Heat deposition via light absorption in nanostructures constitutes a useful tool for controlling nanoscale thermal sources \cite{paper147,B17,paper294}, with potential application in photothermal therapy \cite{GLH07,JEE07,ZK09}, nanoimaging \cite{BTM02,BLB06}, nanocatalysis \cite{PH01}, data storage \cite{WL08}, and hot-electron devices \cite{C14}. Importantly, plasmons in metallic nanostructures enable resonant enhancement of photothermal effects, which can be manipulated down to the nanometer scale \cite{paper147,B17,paper294}. Recently, highly doped graphene has emerged as an outstanding two-dimensional material capable of supporting extremely confined surface plasmons that can be actively tuned by varying its Fermi energy through electrical gating and chemical doping \cite{WSS06,BF07,HD07,JBS09,FAB11,JGH11,paper176,paper196,FRA12,YLC12,YLL12,BJS13,YAL14,paper235}, with application in optical modulation \cite{paper182,AFR12,BJS13,JBS14,KJB18}, light detection \cite{JGH11,YLC12,FLZ13,paper275,LGW16,LLL16}, and sensing \cite{paper256,paper279,HYZ16}. Additionally, the photothermal response of graphene is particularly appealing because of the combination of the following three properties: (i) the low number of electrons needed to sustain plasmons in this material compared with conventional three-dimensional metallic structures; (ii) its low electronic heat capacity; and (iii) the strong variation of its optical response produced by electronic heating. Properties (i) and (ii) lead to unusually high electron temperatures under resonant pumping conditions \cite{paper235,NWG16,paper286}, while properties (ii) and (iii), which originate in the conical electronic bands of graphene \cite{paper235}, give rise to an extraordinary photothermal response.

Due to the generally weak electron-phonon coupling in graphene, the electron temperature can reach significantly high values above the lattice background \cite{GPM13,NWG16}. However, electron-phonon coupling has a dramatic dependence on material quality: in exfoliated clean samples, it is extremely weak and characterized by a relaxation rate that depends linearly on electron-temperature; in contrast, more disordered CVD graphene samples are characterized by a higher relaxation rate with a cubic dependence on electron temperature. The possibility of spatially patterning the electron temperature to manipulate the optical response of graphene appears as a potentially useful approach that deserves further investigation.

In this Letter, we theoretically investigate the photothermally induced optical response of graphene and reveal a radically different behavior in clean and disordered layers leading to unprecedented plasmonic behavior. More precisely, we account in a self-consistent manner for the interplay between optical absorption, heat dissipation, and spatial modification of the electron temperature and optical conductivity under attainable continuous-wave (CW) illumination, and find that weak electron-phonon coupling in clean graphene results in high electron temperatures, while the lattice stays near the ambient level. We exploit this effect to predict (1) a dramatic photothermal modulation of plasmons in graphene ribbons and (2) the existence of plasmons that couple efficiently to external light in homogeneous extended graphene by photothermally patterning a periodic modulation of the optical response. These results illustrate the potential of photothermal engineering to control the plasmonic properties of both structured and extended graphene.

\section{Results}

We adopt the local random-phase approximation (local-RPA) to describe the temperature-dependent optical conductivity of graphene $\sigma(\omega)$ \cite{GSC09,paper235,paper286,paper303}, combined with a two-temperature model \cite{GMB03} to characterize the position-dependent electron and lattice temperatures ($\Te$ and $\Tl$) under CW illumination in the steady-state regime of heat dissipation (see details in the Appendix). The model incorporates the 2D in-plane thermal electron and lattice conductivities ($\ke$ and $\kl$, obtained from their bulk counterparts by multiplying by the graphene thickness $t=0.33\,$nm) to self-consistently calculate the temperature spatial distributions, which are imprinted on the optical conductivity $\sigma$ through its dependence on $\Te$ \cite{GSC09,paper235,paper286,paper303}. Electron-phonon coupling is accounted for by a power-density coupling $g\left(\Te-\Tl\right)$ for clean graphene and $A\left(\Te^3-\Tl^3\right)$ for disordered graphene \cite{VH10,SRL12,BJP13,GSR13,MPR16}, where $g$ and $A$ are material-quality-dependent constant coefficients. Additionally, we phenomenologically introduce thermal coupling from the graphene lattice to the substrate through a term $G\left(\Tl-T_0\right)$, where $G$ is a thermal boundary conductance and $T_0$ is the ambient temperature. In our simulations, we take $T_0=300\,$K and assume parameter values consistent with reported measurements (see Appendix for further details): $g\sim10^{4}\,\mathrm{W/m^{2}K}$, $A=2.24\,\mathrm{W/m^{2}K^{3}}$, and $G=5\,\mathrm{MW/m^{2}K}$ \cite{LPK12,FLA13,THP17}; $\kl/t=100\,$W/mK \cite{YTI13,CSW16}; and $\ke=0.1\kl$ \cite{KPM16}. Specific values for the Fermi energy $\EF$ and the $\EF$-dependent coefficient $g$ are given in the figure captions. Regarding optical damping, we assume a conservative inelastic scattering time $\tau=66\,$fs ($\hbar\tau^{-1}=10\,$meV) in both clean and disordered graphene. Although higher $\tau$'s have been observed in clean graphene at low temperatures \cite{DSB08,WMH13}, it varies with temperature \cite{JKW16} and our assumed value is realistic when considering high electron temperatures. We use a finite-element method for the latter and iterate the electromagnetic and thermal solutions until self-consistency is achieved typically after $\sim10$ iterations. We consider graphene either supported or embedded in an isotropic dielectric of permittivity $\eh=4.4$.

\begin{figure}
\begin{centering}
\includegraphics[width=0.4\textwidth]{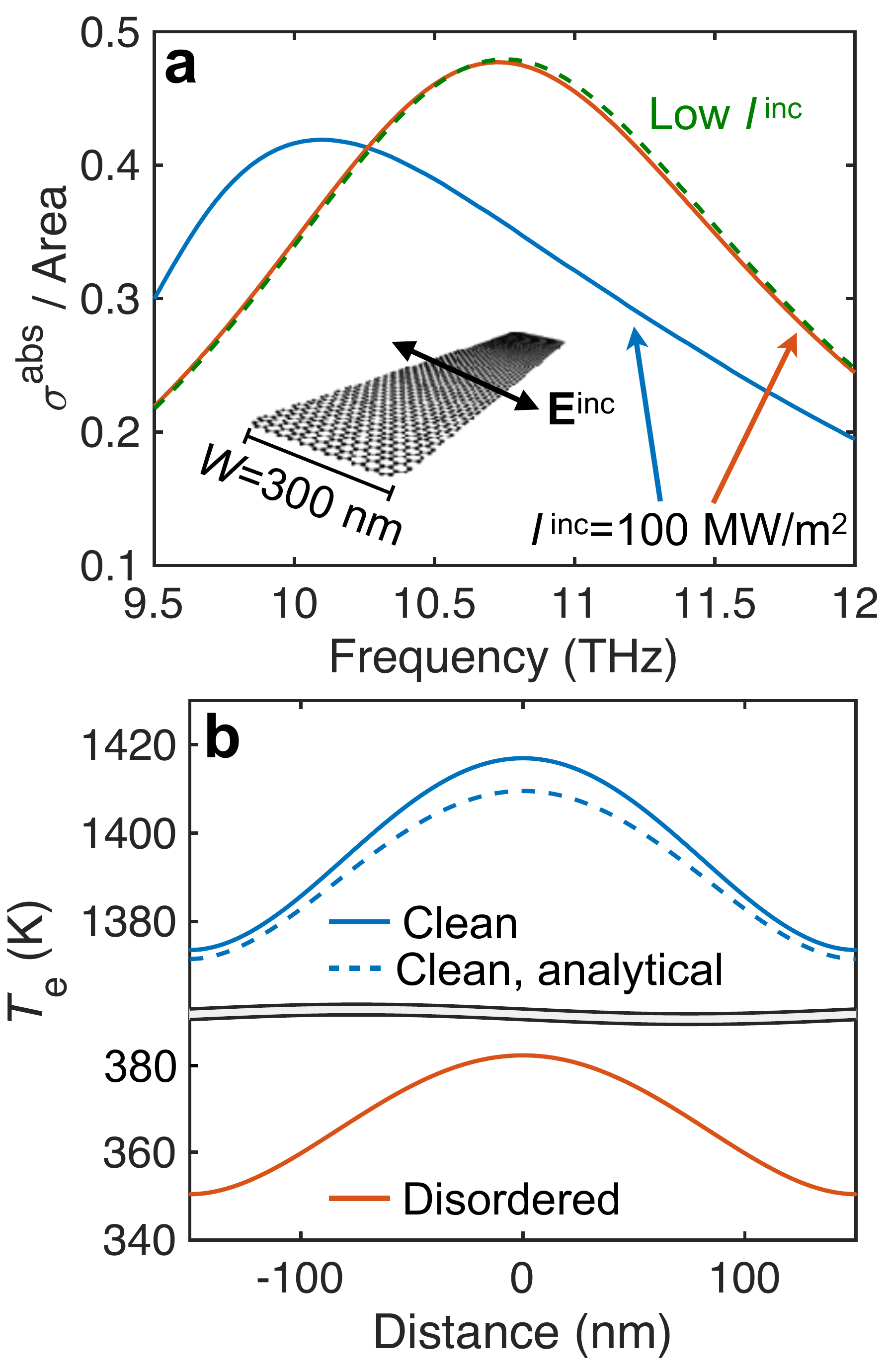}
\par\end{centering}
\caption{Plasmon photothermal effect in a graphene ribbon. (a) Normalized normal-incidence absorption cross-section spectra at the spectral peak for a nanoribbon (width $W=300\,$nm, $\EF=0.4\,$eV, $g=3.84\times10^{4}\,\mathrm{W/m^{2}K}$, embedded in $\eh=4.4$) made of either clean or disordered graphene (solid curves), based on a self-consistent description of heat dissipation for an incident light intensity $\Iinc=100\,\mathrm{MW/m^{2}}$. We show results in the low $\Iinc$ limit for comparison (dashed curve, obtained analytically \cite{paper228,paper293,paper303}, see the Appendix). (b) Variation of the electron temperature across the ribbon for clean and disordered graphene. We find the lattice temperature to be close to the assumed ambient value of $300\,$K (see Appendix). We describe graphene using the temperature-dependent local-RPA conductivity \cite{paper235} with a phenomenological inelastic lifetime $\tau=66\,$fs ($\hbar\tau^{-1}=10\,$meV). Thermal parameters are given in the main text.} \label{Fig1}
\end{figure}

We first study a graphene ribbon (width $W=300\,$nm) under normal-incidence illumination with transversal polarization [inset to Fig.\ \ref{Fig1}(a)]. A prominent plasmon is observed in the absorption spectrum of Fig.\ \ref{Fig1}(a) for low light intensity (dashed curve). The spectrum remains nearly unchanged at high intensity ($\Iinc=100\,\mathrm{MW/m^{2}}$) in disordered graphene (red solid curve), whereas the plasmon peak undergoes a $\sim10\%$ redshift in clean graphene (blue solid curve). We attribute this different behavior to the much weaker electron-phonon coupling in clean graphene \cite{MPR16}, which leads to an elevated electron temperature $\Te\sim1400$\,K, in stark contrast to the mild increase in $\Te$ for disorder graphene, as shown in Fig.\ \ref{Fig1}(b).

\begin{figure}
\begin{centering}
\includegraphics[width=0.4\textwidth]{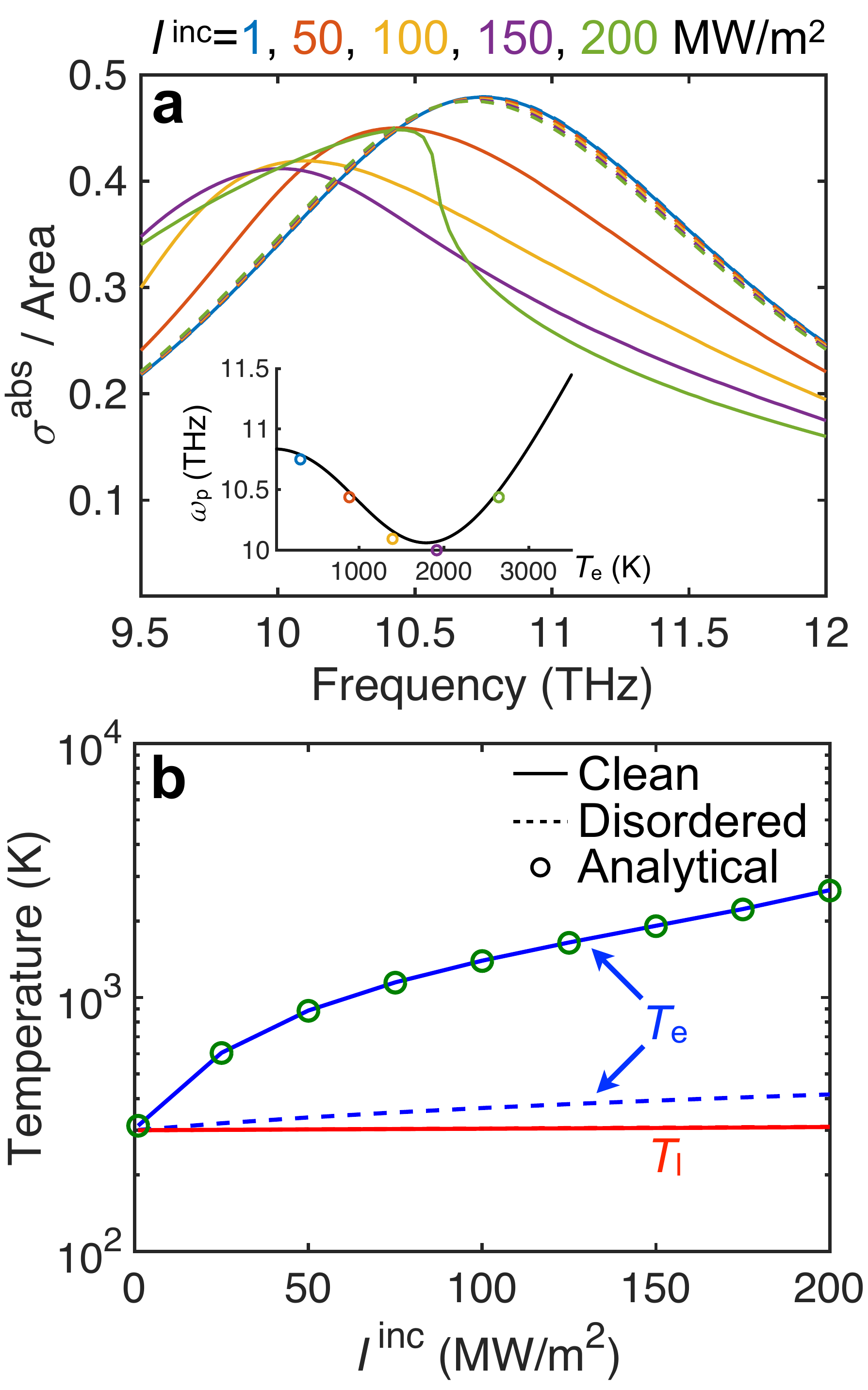}
\par\end{centering}
\caption{Light-intensity dependence of the plasmon photothermal effect. (a) Absorption spectra of the graphene nanoribbon considered in Fig.\ \ref{Fig1} for a wide selection of light intensities. The spectra are dominated by the lowest-order transverse plasmon, the frequency of which is shown in the inset as a function of electron temperature. (b) Space-averaged temperatures (lattice $\Tl$ and electrons $\Te$) as a function of light intensity at the absorption peak frequencies of (a). We present results for both clean (solid curves) and disordered (dashed curves) graphene.} \label{Fig2}
\end{figure}

When varying the incident light intensity in the $\Iinc=$1-100\,$\mathrm{MW/m^{2}}$ range, we find a systematic redshift and broadening of the absorption peak in the clean graphene ribbon (Fig.\ \ref{Fig2}). Further increase in intensity up to $200\,\mathrm{MW/m^{2}}$ produces a large distortion in the absorption spectrum, resulting from the non-monotonic temperature dependence of both the graphene conductivity and the resulting transverse ribbon plasmon energy  (see Fig.\ \ref{Fig4} in the Appendix). The latter admits the analytical expression \cite{paper235} \[\hbar\wp=(e/\sqrt{-\pi\eta_1\eh})\sqrt{\mu^{\rm D}/W},\] where $\eta_1=-0.0687$ is an eigenvalue corresponding to the lowest-order dipolar transverse plasmon, while \[\mu^{\rm D}=\mu+2\kB \Te \log\left(1+e^{-\mu/\kB\Te}\right)\] and \[\mu=\EF\left[\left(1+\xi^4\right)^{1/2}-\xi^2\right]^{1/2},\] with $\xi=(2\log^24)(\kB\Te/\EF)^2$, are the temperature-corrected Drude weight and chemical potential, respectively \cite{paper286}. This expression [solid curve in the inset to Fig.\ \ref{Fig2}(a)] is in excellent agreement with the computed spectral peaks (symbols) when using the calculated spaced-averaged values of $\Te$ as input. Interestingly, the plasmon FWHM is smaller for $\Iinc=200\,\mathrm{MW/m^{2}}$ than for dimmed illumination (Figs.\ \ref{Fig1} and \ref{Fig5} in the Appendix).

Remarkably, under these attainable conditions, the electrons reach a temperature above 2500\,K in clean graphene, while the lattice remains near the ambient level [Fig.\ \ref{Fig2}(b)]. We stress again that this is in stark contrast to disordered graphene, for which the spectra remain nearly unchanged within the considered intensity range and the electron temperature hardly exceeds 400\,K (Fig.\ \ref{Fig2}(b) and Fig.\ \ref{Fig6} in the Appendix) due to a more efficient electron-phonon coupling.

We obtain further insight into the photothermal response of clean graphene by adopting the reasonable assumption $\Tl\approx T_0$, which effectively decouples the lattice from the electronic system (see Figs.\ \ref{Fig6}-\ref{Fig8} in the Appendix), so that heat dissipation is fully described through
\begin{align}
\nabla\cdot\ke\nabla(\Te-T_0)-g(\Te-T_0)\approx-p^{\rm abs},
\label{heat}
\end{align}
where $p^{\rm abs}$ is the power density of optical absorption. Further assuming a constant value of $\ke$, this equation allows us to obtain a characteristic electronic-heat-diffusion distance $D_{\rm e}=\sqrt{\ke/g}$. Indeed, a measure of the degree of heat localization is provided by the electron temperature profile produced by a line heat source, $\Te(x)\propto\ee^{-|x|/D_{\rm e}}$, as a function of distance $x$ to it. Under the conditions of Figs.\ \ref{Fig1} and \ref{Fig2}, we have $D_{\rm e}\approx293\,$nm$\sim W$, which explains why $\Te$ is nearly uniform across the ribbon, unlike the cosine-like $p^{\rm abs}$ transversal profile associated with the dipolar plasmon under consideration (see Fig.\ \ref{Fig9} in the Appendix). The uniformity of $\Te$ now allows us to write the analytical estimate \[\Te=T_0+\Iinc(\sigma^{\rm abs}/{\rm Area})/g\] for clean graphene, represented in Fig.\ \ref{Fig2}(b) [symbols, with $\sigma^{\rm abs}/{\rm Area}$ taken at the peak frequencies of Fig.\ \ref{Fig2}(a)], in excellent agreement with full numerical simulations (solid blue curve). 

Incidentally, the values of the relaxation time $\tau$ and the out-of-plane thermal conductance $G$, which depend on both the material quality and the properties of the surrounding media, affect $\Te$ and $\Tl$: they increase with decreasing $G$, while the opposite behavior is observed for decreasing optical damping $\tau^{-1}$ (see Fig.\ \ref{Fig7} in the Appendix). Nevertheless, the influence of $G$ is minor under the conditions here considered because $\Tl\approx T_0$.

Plasmon confinement in graphene has so far been achieved through lateral patterning (e.g., in ribbons \cite{BJS13,JBS14}), inhomogeneous doping \cite{ROG16}, or nanostructured dielectric environments \cite{GSJ13,ZYU13}. These approaches require the use of nanolithography, which is generally detrimental for the graphene quality. Motivated by the above study for graphene ribbons, we explore next a radically different method for producing and actively tuning plasmon confinement in extended, unpatterned graphene that does not require nanostructuring: spatially modulated optical heating can be applied by projecting an on-demand pump pattern, thus configuring an inhomogeneous graphene optical response capable of trapping plasmons and molding their spatial profiles with a resolution limited by far-field diffraction to roughly half the pump wavelength $\lambda_{\rm pump}/2$.

\begin{figure}
\begin{centering}
\includegraphics[width=0.4\textwidth]{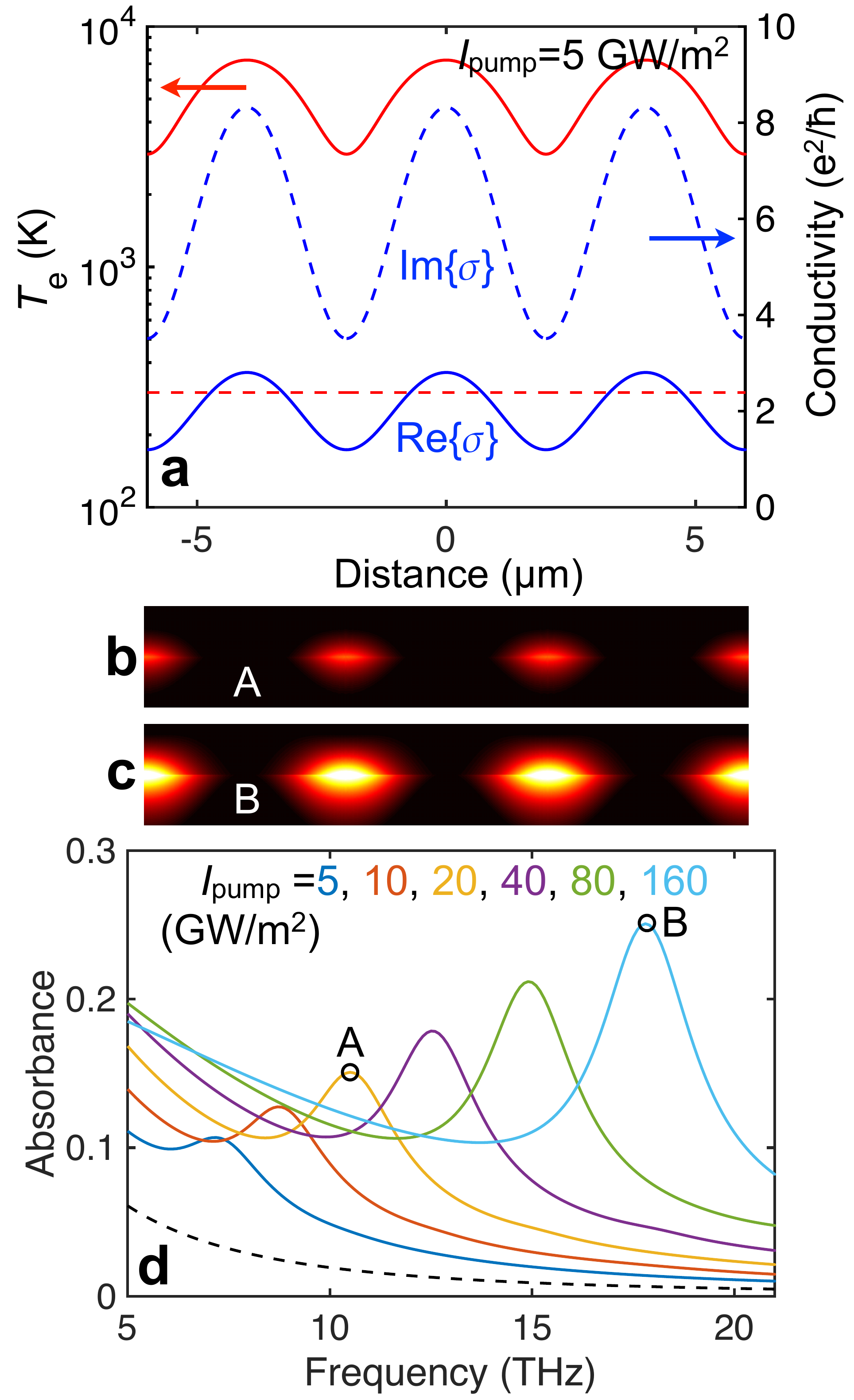}
\par\end{centering}
\caption{Photothermal patterning of the optical response of homogeneous graphene. We consider an extended doped clean graphene sheet ($\EF=0.3\,$eV, $g=1.21\times10^{4}\,\mathrm{W/m^{2}K}$, substrate $\eh=4.4$) exposed to a light intensity grating of $4\,\mu$m period, formed by the interference of two CW 785\,nm pump plane waves of equal intensity ($I_{\rm pump}$ each). (a) Spatial variation of the self-consistent electron temperature produced by the pump with $I_{\rm pump}=5\,\mathrm{GW/m^2}$ (red solid curve, left scale) and resulting optical conductivity in the local-RPA model (blue curves, right scale) at a probing frequency of 7.2\,THz. The ambient temperature level (300\,K) is shown for reference (red broken line). (b,c) Near-field intensity plots in a plane transversal to the graphene at the peak plasmon frequencies shown in (d) for $I_{\rm pump}=20\,\mathrm{GW/m^2}$ (A) and $160\,\mathrm{GW/m^2}$ (B). (d) Absorption spectra probed in the THz region with (solid curves) and without (dashed curve) light grating pumping for various values of $I_{\rm pump}$.} \label{Fig3}
\end{figure}

We demonstrate the feasibility of this concept by considering an extended clean graphene sheet ($\EF=0.3\,$eV doping, supported on a substrate $\eh$), on which a pump {\it light grating} is formed by interfering two coherent s-polarized CW plane waves ($\lambda_{\mathrm {pump}}=785\,$nm, intensity $I_{\mathrm {pump}}=5-160\,\mathrm{GW/m^2}$ each, incidence angles $\pm\theta=\pm5.6^{\circ}$). The in-plane electric-field pump intensity is then $4I_{\mathrm {pump}}\cos^2\left[2\pi\sin\theta (x/\lambda_{\mathrm {pump}})\right]$, where we take the beam directions to lie on the plane formed by the surface normal and the in-plane $x$ axis. Incidentally, we obtain a graphene absorbance $(4\pi{\rm Im}\{\sigma\}/c)|t_{\rm s}|^2\sim0.002-0.008$ from the local-RPA conductivity $\sigma$, with $t_{\rm s}\approx2/(1+\sqrt{\eh})$; this results deviates from the $\Te=0$ result \cite{MSW08,NBG08} $\approx0.023\,t_{\rm s}^2$ due to thermally-driven saturable absorption over the range of $I_{\mathrm {pump}}$ under consideration (see Figs.\ \ref{Fig10} and \ref{Fig12} in the Appendix). The resulting self-consistent electron temperature $\Te$ reaches high values ($\sim7200\,$K for $5\,\mathrm{GW/m^2}$) and displays a periodic pattern with a max-to-min contrast ratio $\sim2.5$ [Fig.\ \ref{Fig3}(a)]. This imparts a periodic pattern on the optical conductivity, effectively transforming the optical response of the extended graphene layer into that of a graphene ribbon array. When examining the absorption spectra as a function of probe frequency in the THz region [Fig.\ \ref{Fig3}(b)], a prominent resonance peak is observed, shifting up in frequency as the pump intensity is increased. Interestingly, the resonant near-field probe intensity distribution [Fig.\ \ref{Fig3}(b), upper inset] reveals plasmon confinement in the minima of $\Te$ regions, where ${\rm Re}\{\sigma\}$ reaches a minimum (i.e., low inelastic losses), while ${\rm Im}\{\sigma\}$ is also minimum and configures an effective plasmon potential well.

Incidentally, in the design of photothermal modulation we need to reduce the electronic-heat diffusion distance below the characteristic pattern distance (e.g., the optical grating period, $\gg D_{\rm e}\approx638\,$nm under the conditions of Fig.\ \ref{Fig3}; see Fig.\ \ref{Fig11} in the Appendix).

\section{Conclusion}

In summary, we have shown that the characteristic weak electron-phonon coupling in clean graphene allows us to reach high electron temperatures well above the lattice temperature, which in turn stays near ambient levels under CW illumination conditions. This produces strong photothermal modulation in the graphene optical response, which we exploit to predict large plasmon shifts in ribbons. We further postulate this effect as an efficient way of dynamically imprinting a spatial modulation of the optical response in extended homogeneous graphene, whereby a spatially patterned optical pump is used to locally heat graphene electrons, thus tailoring an on-demand nanoscale response. We illustrate this concept by showing resonant absorption in a photothermally imprinted grating, whose plasmons can couple to far-field radiation, unlike those of homogenous graphene. Besides circumventing the requirement of nanostructuring, this approach can potentially enable fast plasmon modulation relying on the ability to shape the light pumping beams. 

\section*{Acknowledgments}

This work has been supported in part by the Spanish MINECO (MAT2017-88492-R and SEV2015-0522), the European Commission (Graphene Flagship 696656), AGAUR (2014-SGR-1400), the Catalan CERCA Program, Fundaci\'o Privada Cellex, and the US National Science Foundation (CAREER Award) and Office of Naval Research.

\begin{widetext}

\appendix

\begin{figure}
\begin{centering}
\includegraphics[width=0.65\textwidth]{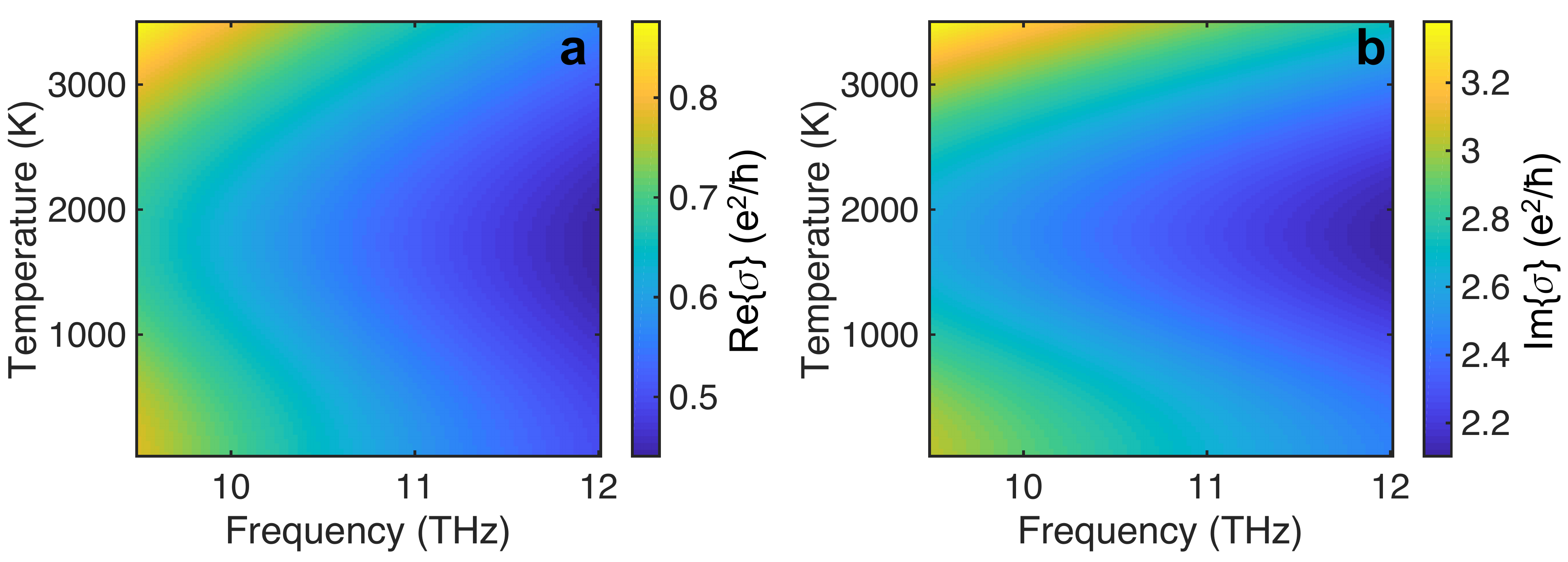}
\par\end{centering}
\caption{Temperature dependence of the optical conductivity of graphene. We plot the real (a) and imaginary (b) parts of the graphene conductivity in the local-RPA limit, calculated from Eq.\ (\ref{eq:sigma}) for $\EF=0.4\,$eV and $\hbar\tau^{-1}=10\,$meV as a function of electron temperature and light frequency.}
\label{Fig4}
\end{figure}

\section{Temperature-dependent graphene chemical potential and optical conductivity}

At zero temperature, the graphene Fermi energy $E_{\mathrm{F}}$ is controlled by the doping carrier density $n$ as $\EF=\hbar v_{\mathrm{F}}\sqrt{\pi n}$, where $v_{\mathrm{F}}\approx10^{6}\,$m/s is the Fermi velocity. At finite electron temperature $T_{\mathrm{e}}$, the occupancy of electronic states is described by the Fermi-Dirac distribution $f_E=\left\{ \exp\left[\left(E-\mu\right)/k_{\mathrm{B}}T_{\mathrm{e}}\right]+1\right\} ^{-1}$, where $E$ is the electron energy and $\mu$ is the chemical potential. The latter is determined by the condition that the number of carriers is conserved, and it can be analytically approximated by the expression \cite{paper286}
\begin{align}
\mu=\sqrt{\sqrt{\left(E_{\mathrm{F}}\right)^{4}+\left(2\log^{2}4\right)^2\left(k_{\mathrm{B}}T_{\mathrm{e}}\right)^{4}}-\left(2\log^{2}4\right)\left(k_{\mathrm{B}}T_{\mathrm{e}}\right)^{2}}.\label{eq:mu}
\end{align}
We characterize the optical properties of graphene through its 2D conductivity, for which we adopt the local limit of the random-phase approximation (local-RPA) \cite{paper235,GSC09}
\begin{align}
\sigma\left(\omega\right)=\frac{e^{2}}{\pi\hbar^{2}}\frac{\mathrm{i}}{\omega+\mathrm{i}\tau^{-1}}\left\{ \mu^{\mathrm{D}}-\int_{0}^{\infty}dE\frac{f_E-f_{-E}}{1-4E^{2}/\left[\hbar^{2}\left(\omega+\mathrm{i}\tau^{-1}\right)^{2}\right]}\right\} ,\label{eq:sigma}
\end{align}
where $\mu^{\mathrm{D}}=\mu+2k_{\mathrm{B}}T_{\mathrm{e}}\ln\left(1+\mathrm{e}^{-\mu/k_{\mathrm{B}}T_{\mathrm{e}}}\right)$
is the effective Drude weight, while $\hbar\tau^{-1}$ is a phenomenological inelastic scattering rate. We plot the temperature-dependent conductivity in Fig.\ \ref{Fig4} for the graphene parameters considered in Figs.\ 1 and 2.

\section{Self-consistent photothermal response of graphene}

We adopt a two-temperature model to characterize the system, which describes the electron and lattice temperatures $\Te$ and $\Tl$ in terms of the equations
\begin{align}
C_{\mathrm{e}}\frac{d\Te}{dt}&=p^{\mathrm{abs}}+\nabla\cdot\left(\ke\nabla\Te\right)-H\left(\Te,\Tl\right),\label{eq:te} \\
C_{\mathrm{l}}\frac{d\Tl}{dt}&=\nabla\cdot\left(\kl\nabla\Tl\right)+H\left(\Te,\Tl\right)-G\left(\Tl-T_{0}\right),\label{eq:tl}
\end{align}
where $T_{0}$ is the ambient temperature, $C_{\mathrm{e}}$ and $C_{\mathrm{l}}$ are the 2D electron and lattice heat capacities (in J/m$^2$\,K), $\ke$ and $\kl$ are the 2D electron and lattice thermal conductivities (in W/K), $p^{\mathrm{abs}}$ is the absorption density (in W/m$^2$), $H\left(\Te,\Tl\right)$ describes electron-phonon coupling (see below), and $G$ is the out-of-plane thermal conductance to the environment. In our simulations, we take $G=5\,\mathrm{MW/m^{2}K}$ when graphene is supported on a substrate \cite{LPK12,FLA13,THP17}, and twice this value it is fully embedded in a homogeneous medium. Under CW illumination, the temperatures reach a steady-state regime, in which the left-hand sides of Eqs.\ (\ref{eq:te}) and (\ref{eq:tl}) vanish, leading to
\begin{align}
-\nabla\cdot\left(\ke\nabla\Te\right)&=p^{\mathrm{abs}}-H\left(\Te,\Tl\right),\label{eq:tes} \\
-\nabla\cdot\left(\kl\nabla\Tl\right)&=H\left(\Te,\Tl\right)-G\left(\Tl-T_{0}\right).\label{eq:tls}
\end{align}
Because $p^{\mathrm{abs}}$ is intrinsically dependent on $\Te$ through the optical conductivity of graphene [see Eq.\,(\ref{eq:sigma})], we need to solve Eqs.\ (\ref{eq:tes}) and (\ref{eq:tls}), together with the optical response (i.e., $p^{\rm abs}$), in a self-consistent manner, for which we use a numerical solver based on finite-element methods (FEM). We assume $T_0=300\,$K throughout this work, unless otherwise stated.

\section{Electron-phonon coupling}

The electron-phonon cooling power density $H\left(\Te,\Tl\right)$ in the limit of large electron mean free path $\ell$, termed clean limit in this work, is known to scale linearly with the difference between $\Te$ and $\Tl$ \cite{MPR16}:
\begin{align}
H\left(\Te,\Tl\right)=g\left(T_{\mathrm{e}}-T_{\mathrm{l}}\right),\label{eq:cleanH}
\end{align}
where
\begin{align}
g=\frac{D^{2}\mu^{4}k_{\mathrm{B}}}{2\pi\rho\hbar^{5}v_{\mathrm{F}}^{6}},\label{eq:g}
\end{align}
$\rho=7.6\times10^{-8}\,\mathrm{g/cm^{2}}$ is the mass density of graphene, and $D$ is the deformation potential of this material. Note that Eq.\,(\ref{eq:cleanH}) is valid for $\EF\gg k_{\mathrm{B}}\Te$ and $\Te\gg T_{\mathrm{BG}}$, where
$T_{\mathrm{BG}}=(2s/\vF)\EF/k_{\mathrm{B}}$ is the Bloch-Gr\"{u}neisen temperature \cite{VH10} ($T_{\mathrm{BG}}\approx186\,$K for $\EF=0.4\,$eV), and $s\approx0.02v_{\mathrm{F}}$ is the speed of sound in graphene.

The presence of disorder has a strong influence on the electron-phonon coupling, producing a departure from a linear to a cubic temperature dependence of the cooling power density \cite{SRL12}:
\begin{align}
H\left(\Te,\Tl\right)=A\left(T_{\mathrm{e}}^{3}-T_{\mathrm{l}}^{3}\right),\label{eq:disorderH}
\end{align}
where
\begin{align}
A=\frac{1.2D^{2}\left|\mu\right|k_{\mathrm{B}}^{3}}{\pi^{2}\rho\hbar^{4}v_{\mathrm{F}}^{3}s^{2}\ell}.\nonumber
\end{align}

In the simulations presented in Figs.\ 1, 2, and \ref{Fig4}-\ref{Fig9}, we assume $\mu\approx0.4\,$eV, $D=20\,$eV (40\,eV) in clean (disordered) graphene, and $\ell=10\,$nm, leading to $g=3.84\times10^{4}\,\mathrm{W/m^{2}K}$ and $A=2.24\,\mathrm{W/m^{2}K^{3}}$. In Figs.\ 3 and \ref{Fig11}, we assume the above value of $D$ for clean graphene and a chemical potential $\mu\approx0.3\,$eV, leading to $g=1.21\times10^{4}\,\mathrm{W/m^{2}K}$.

\section{Electron and lattice thermal conductivities in graphene}

Although the electronic thermal conductivity of graphene $\ke$ depends on both the carrier density $n$ and the electron temperature $\Te$ \cite{YTI13,CSW16}, we assume a constant value with a similar order of magnitude as measured in experiment for the temperature range considered in this work. More precisely, we take a 3D conductivity $\ke/t=10\,$W/m\,K, where $t=0.33\,$nm is the graphene thickness. Additionally, we set the lattice thermal conductivity to $\kl/t=100\,$W/m\,K, which is also consistent with experiment \cite{KPM16}.

\section{Analytical solution to the ribbon absorption at low intensity [dashed curve in Fig.\ 1(a)]}

The smallness of the plasmon wavelength compared with the light wavelength and the two-dimensionality of graphene have been extensively employed to produce accurate analytical solutions for the plasmonic response of different graphene structures, including graphene ribbons \cite{paper235,paper303}. The optical extinction and absorption cross-sections are then nearly identical, and for a ribbon embedded in a homogeneous medium of permittivity $\eh$ we have \cite{paper235,paper303}
\begin{align}
\frac{\sigma^{\mathrm{abs}}\left(\omega\right)}{A}=\frac{4\pi\omega}{\sqrt{\eh}c}\,\mathrm{Im}\left\{\frac{\alpha_\omega}{A}\right\},
\end{align}
where $A$ is the graphene area, 
\begin{align}
\frac{\alpha_\omega}{A}= -\eh W\sum_j\frac{\zeta_j^2}{\ii\eh\omega W/\sigma(\omega)+1/\eta_j}\label{eq:alpha}
\end{align}
is the polarizability polarizability per unit area, and $\eta_j$ and $\zeta_j$ and eigenvalues and dipoles associated with different plasmon modes $j$. This expression is valid for normal incidence and polarization across the ribbon. In Fig.\ 1(a), we evaluate the above expression considering only the dominant lowest-order dipolar mode, for which we use \cite{paper303} $\eta_1=-0.0687$ and $\zeta_1=0.943$.

\section{Analytical solution to the electron temperature in a ribbon [dashed curve in Fig.\ 1(b)]}

In the clean graphene limit, $H\left(\Te,\Tl\right)$ is given by Eq.\,(\ref{eq:cleanH}) and we empirically find $\Tl\approx T_{0}$ [see Figs.\ \ref{Fig6}(b) and \ref{Fig10}(c,f)]. We can then impose $\Tl=T_0$ and further approximate the absorption density as $p^{\mathrm{abs}}=P\cos\left(\pi x/W\right)$, where $P=(\pi/2)I^{\rm inc}\sigma^{\rm abs}/A$ is adjusted in order to reproduce the total absorbed power $I^{\rm inc}\sigma^{\rm abs}$ (see Fig.\ \ref{Fig9}), while $x$ is the spatial coordinate across the ribbon (with $-W/2<x<W/2$). This approximation allows us to solve Eq.\ (\ref{eq:tes}) analytically:
\begin{align}
\Te=T_{0}+\frac{P\cos\left(\pi x/W\right)}{\pi^{2}\ke/W^{2}+g}+C\left[\exp\left(\frac{x+W/2}{D_{\mathrm{e}}}\right)+\exp\left(-\frac{x-W/2}{D_{\mathrm{e}}}\right)\right],\label{eq:tecana}
\end{align}
where $D_{\mathrm{e}}=\sqrt{\ke/g}$ is a characteristic thermal diffusion distance and $C$ is an integration constant that we fit in such a way that the electron-phonon cooling rate fully compensates for optical absorption [i.e., $I^{\rm inc}\sigma^{\rm abs}/A=(g/W)\int^{W/2}_{-W/2}\left(\Te-T_0\right)dx$]. This procedure, which we can apply to any choice of geometrical and illumination parameters as long as we can approximate $\Tl\approx T_{0}$, produces the dashed curve in Fig.\,1(b), in good agreement with the numerically calculated $\Te$ profile, except for a small underestimate of $\Te$, essentially accounted by the nearly uniform difference $\Tl-T_0\approx4\,$K obtained in the self-consistent numerical calculation.

\section{Additional simulations}

We present additional simulations in Figs.\ \ref{Fig5}-\ref{Fig11} (see main text for further discussion).

\begin{figure}
\begin{centering}
\includegraphics[width=1\textwidth]{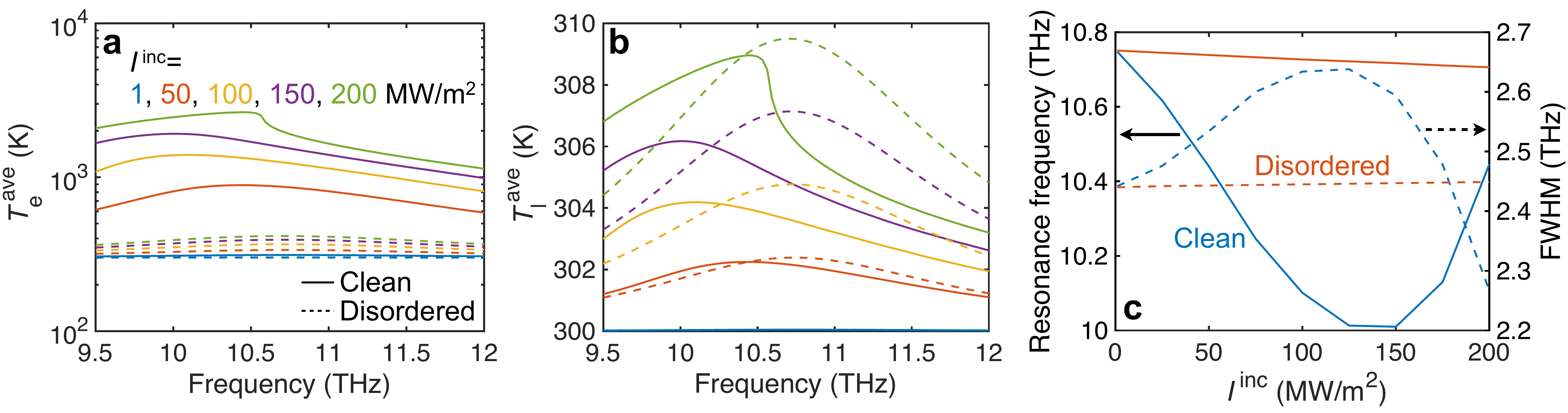}[H]
\par\end{centering}
\caption{Intensity dependence of optical heating in the graphene ribbon of Fig.\ 1. (a,b) Surface-averaged electron (a) and lattice (b) temperatures as a function of light frequency for various incident light intensities [see labels in (a)] using either clean (solid curves) or disordered (broken curves) graphene. (c) Resonance frequency (left vertical axis) and FWHM (right axis) extracted from the absorption spectra in Fig.\ 2(a) as a function of light intensity.}
\label{Fig5}
\end{figure}

\begin{figure}
\begin{centering}
\includegraphics[width=0.95\textwidth]{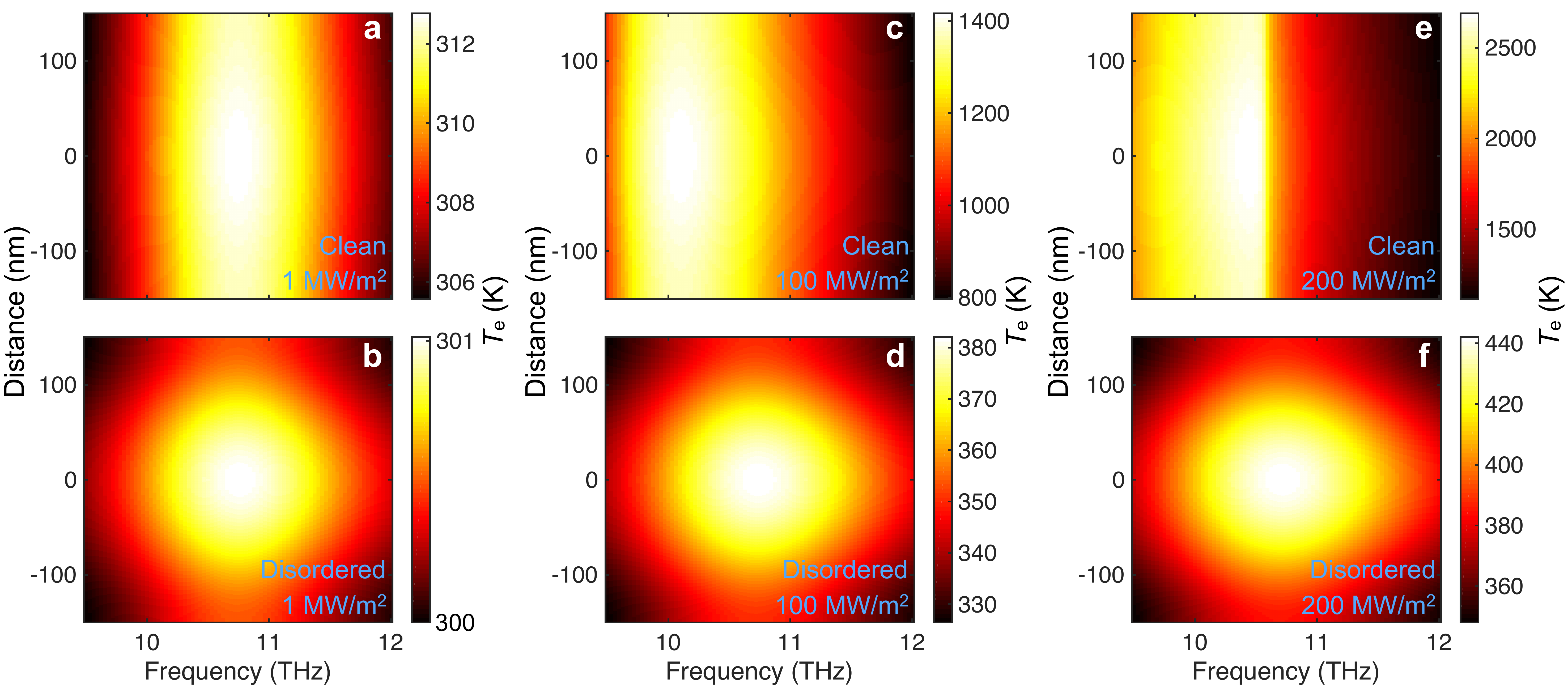}
\par\end{centering}
\caption{Variation of the electron temperature landscape with incident light intensity. We plot the electron temperature as a function of distance across the ribbon of Fig.\ 1 and incident light frequency for clean and disordered graphene. Various values of the light intensity are considered (see labels).}
\label{Fig6}
\end{figure}

\begin{figure}
\begin{centering}
\includegraphics[width=0.95\textwidth]{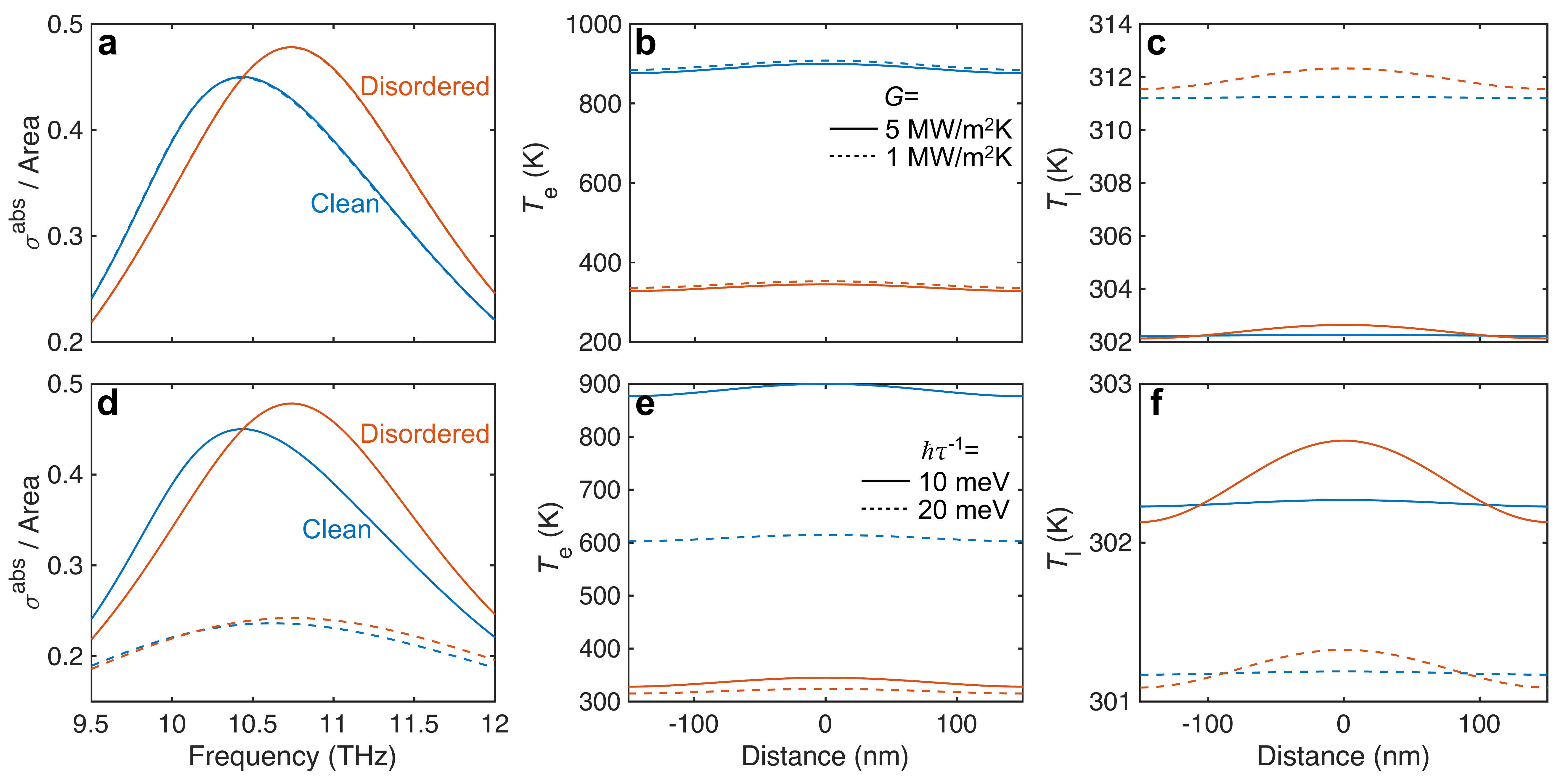}
\par\end{centering}
\caption{Dependence of the photothermal plasmonic response on out-of-plane thermal conductance $G$ and inelastic scattering rate $\tau^{-1}$. We present absorption spectra (a,d), as well as electron (b,e) and lattice (c,f) spatial temperature distributions [calculated at the frequencies of the absorption peaks in (a,b)], for the same graphene nanoribbon considered in Fig.\,1 with an incident light intensity of $50\,\mathrm{MW/m^{2}}$. (a-c): dependence on the out-of-plane thermal conductance $G$. (d-f): dependence on the inelastic scattering rate $\tau^{-1}$. We consider both clean (blue curves) and disordered (red curves) graphene. }
\label{Fig7}
\end{figure}

\begin{figure}
\begin{centering}
\includegraphics[width=0.65\textwidth]{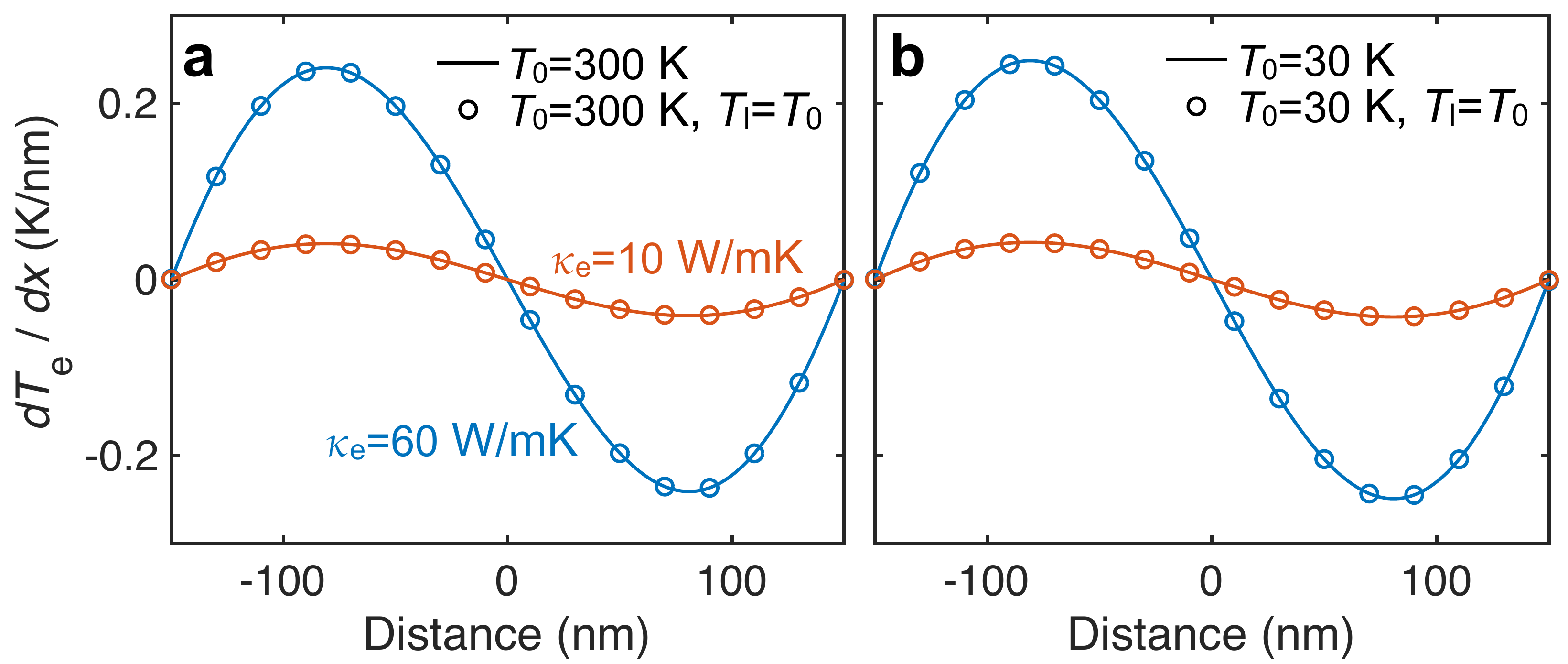}
\par\end{centering}
\caption{Temperature gradient in the ribbon of Fig.\ 1 for different background temperatures with (symbols) and without (solid curves) the assumption of $\Tl=T_0$. We consider two different values of the electronic thermal conductivity (see labels). The light intensity is $I^{\mathrm{inc}}=50\,\mathrm{MW/m^{2}}$. The lattice conductivity is taken as $\kl=10\ke$ in all cases.}
\label{Fig8}
\end{figure}

\begin{figure}
\begin{centering}
\includegraphics[width=0.4\textwidth]{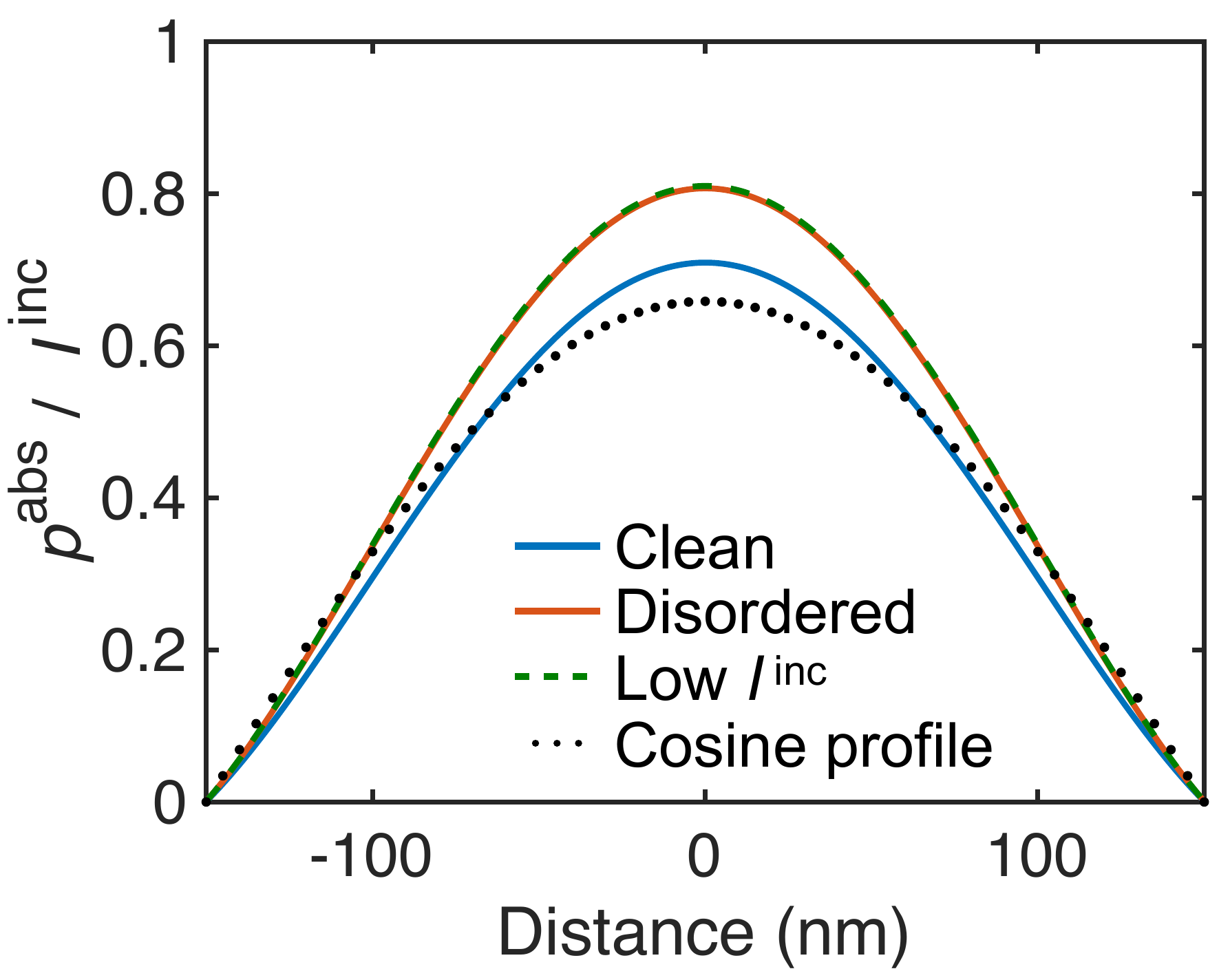}
\par\end{centering}
\caption{Normalized absorption density at the spectral peak of the ribbon considered in Fig.\ 1. We consider both clean and disordered graphene. We also plot the analytical result derived from a plasmon-wave-function analysis in the low-intensity limit \cite{paper303} (dashed curve). We further compare with a cosine-like profile (dotted curve) that has the same area as the absorption density for clean graphene (i.e., the same total absorbed power).}
\label{Fig9}
\end{figure}

\begin{figure}
\begin{centering}
\includegraphics[width=0.75\textwidth]{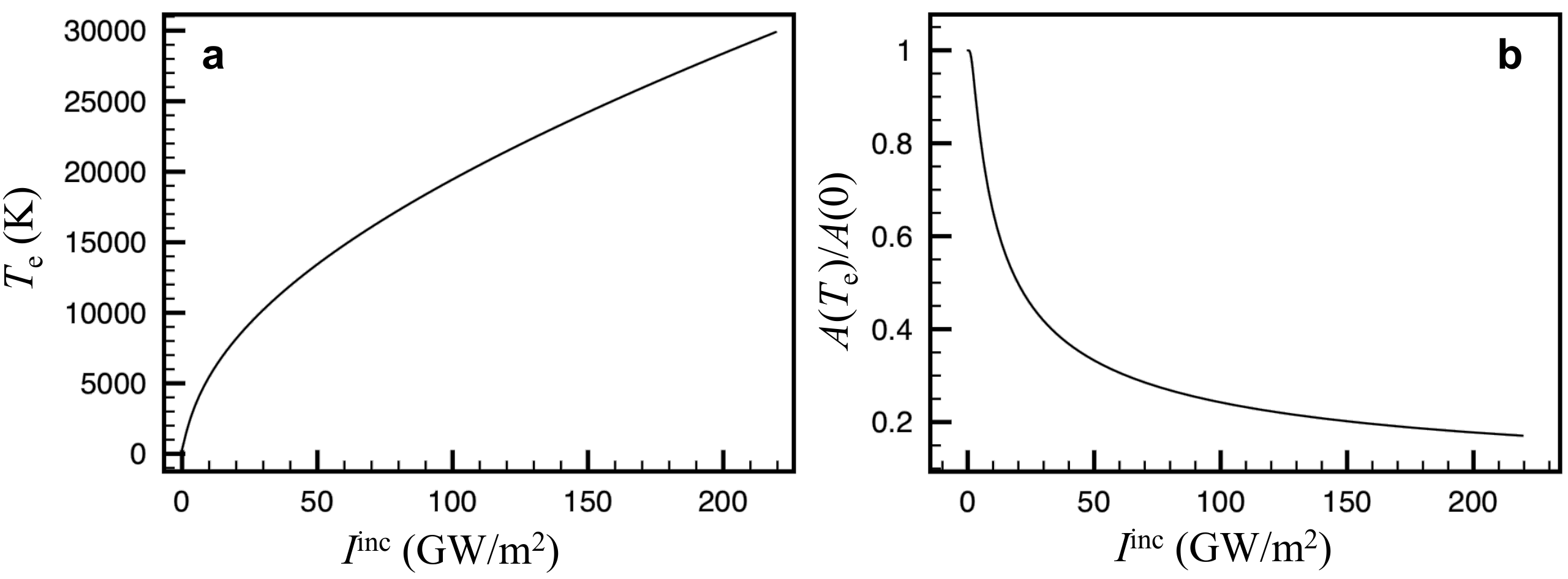}
\par\end{centering}
\caption{Dependence of the average electron temperature (a) and the absorbance (b) as a function of light intensity in extended homogeneous graphene. We present the normal-incidence absorbance $\mathcal{A}=(4\pi{\rm Re}\{\sigma\}/c)|t_{\rm s}|^2$ calculated for graphene on a substrate of permittivity $\eh=4.4$. Here, $t_{\rm s}\approx2/(1+\sqrt{\eh})$ is the Fresnel transmission coefficient and $\sigma$ is the RPA graphene conductivity [notice that the local limit of Eq.\ (\ref{eq:sigma}) is exact under normal incidence]. The latter depends on the electron temperature $\Te$, which we evaluate self-consistently through the approximation $\Te=T_0+I^{\rm inc}\mathcal{A}/g$ (see main text). The graphene parameters are the same as in Fig.\ 3 and the light wavelength is 785\,nm. Although these results are obtained for unstructured, normally-incident light, they allow us to visualize the behavior of $\Te$ and $\mathcal{A}$ as a function of light intensity for application to photothermal engineering of extended graphene.}
\label{Fig10}
\end{figure}

\begin{figure}
\begin{centering}
\includegraphics[width=0.50\textwidth]{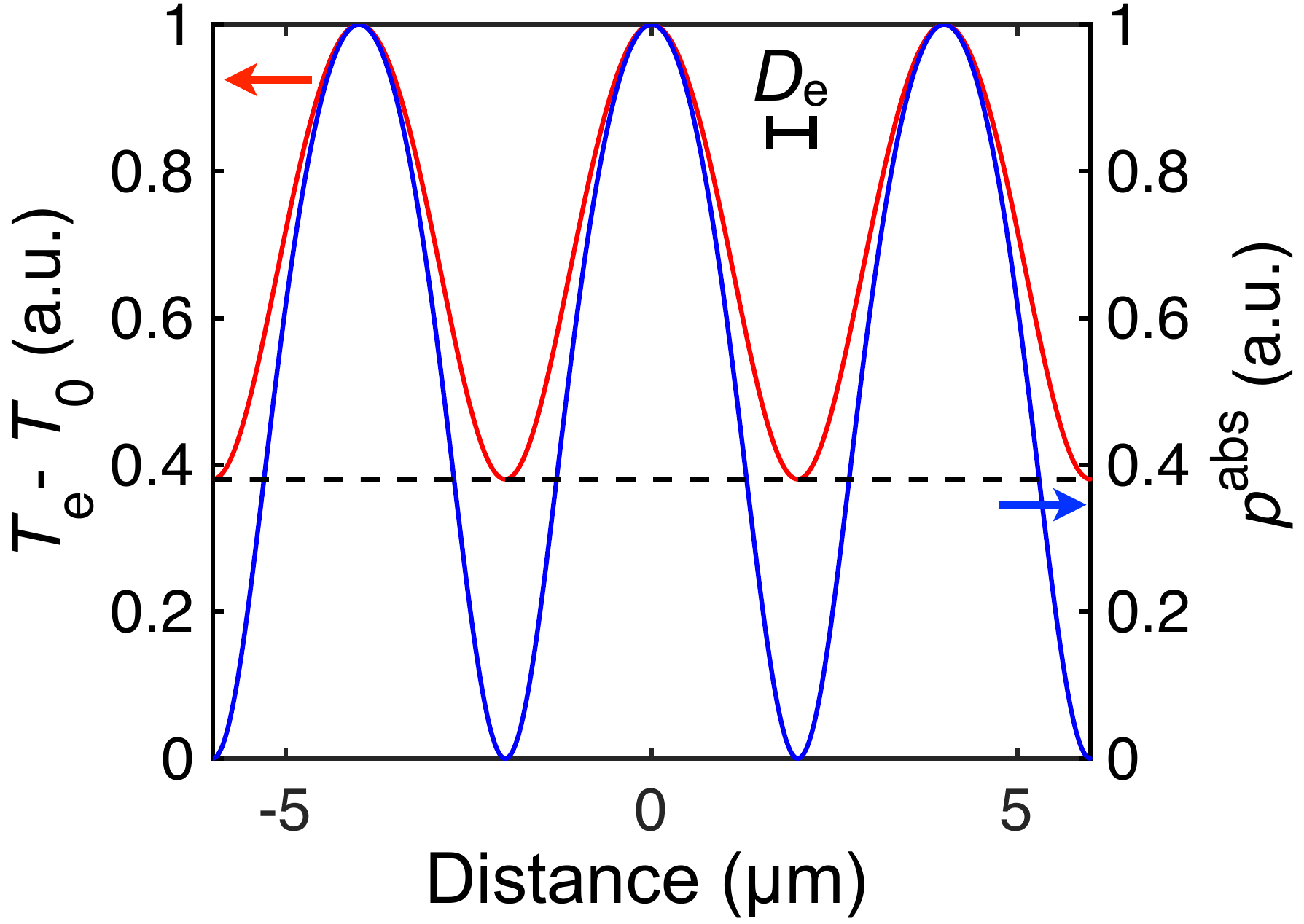}
\par\end{centering}
\caption{Heat diffusion in extended graphene under photothermal patterning. We plot the electron temperature (red curve, left scale) and the self-consistently calculated absorption density (blue curve, right scale) as a function of distance along the plane of extended graphene under the condietions considered in Fig.\ 3(a). Both curves are normalized to their respective maximum values. The scale bar shows the characteristic diffusion length $D_{\mathrm{e}}=\sqrt{\ke/g}$, in good qualitative agreement with the broadening of the temperature distribution compared with the absorption density.}
\label{Fig11}
\end{figure}

\begin{figure}
\begin{centering}
\includegraphics[width=0.75\textwidth]{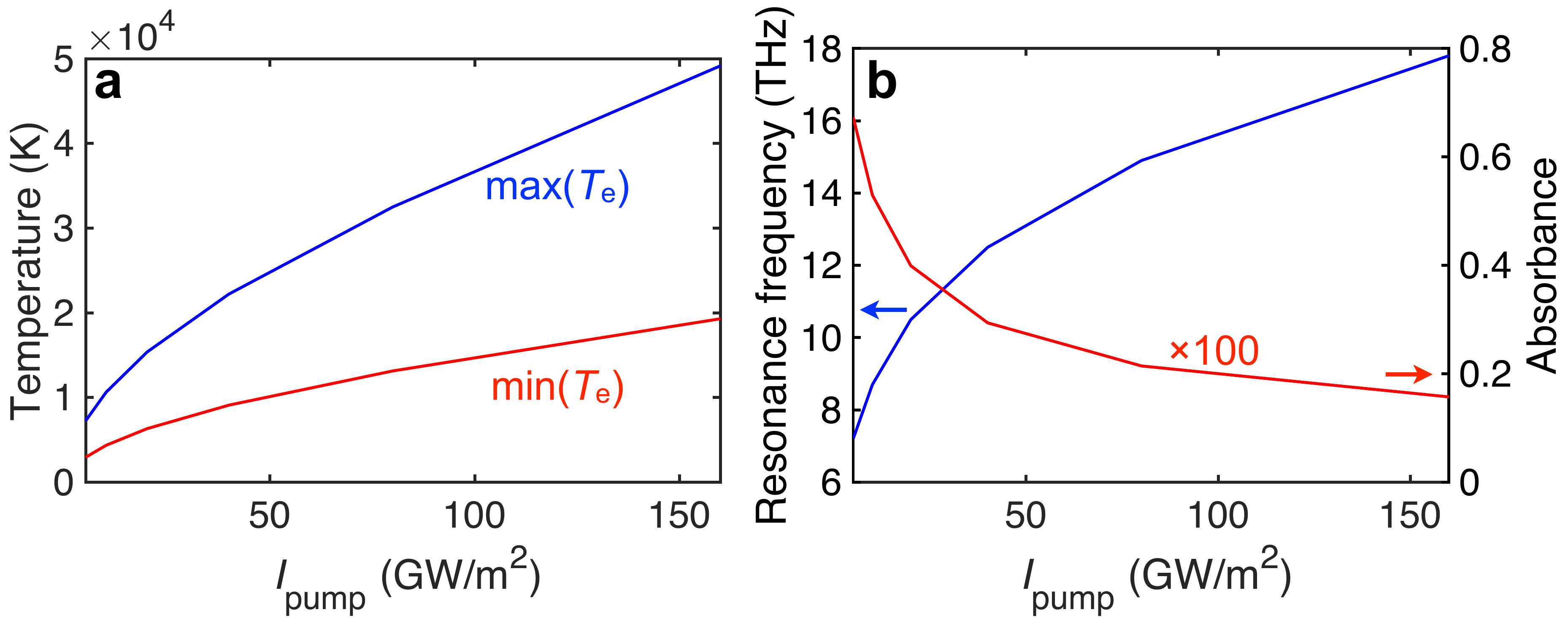}
\par\end{centering}
\caption{Dependence of various parameters characterizing the photothermal patterning of Figs.\ 3 and \ref{Fig11} as a function of the pump intensity $I_{\rm pump}$ of each of the two interfering beams. (a) Maximum and minimum temperatures produced by the optical grating. (b) Normal-incidence plasmon frequency (left scale) and absorbance at the pump wavelength (right scale, curve multiplied by 100).}
\label{Fig12}
\end{figure}

\end{widetext}


\end{document}